\documentclass[12pt]{iopart}

\usepackage{iopams}
\usepackage{cite}
\usepackage{float}
\usepackage{graphicx}
\usepackage{epstopdf}
\usepackage{xcolor}
\usepackage{caption}
\usepackage{subfigure}

\begin{document}

\title{Multiple Pulses Phase-matching Quantum Key Distribution}

\author{Gang Chen$^{1**}$, Le Wang$^{1**}$, Wei Li$^1$, Qianping Mao$^1$, Zhigang Shen$^1$, Shengmei Zhao$^{1,2*}$, and Jozef Gruska$^3$}

\address{1 Institute of Signal Processing Transmission, Nanjing University of Posts and Telecommunications (NUPT), Nanjing, 210003,  China}
\address{2 Key Lab of Broadband Wireless Communication and Sensor Network Technology, Ministry of Education, Nanjing, 210003, China}
\address{3 Faculty of informatics, Masaryk University, Botanick\'{a} 68a, Brno, 60200, Czech Republic}
\address{ *Author to whom any correspondence should be addressed}
\address{ **These authors contributed equally to this work}
\ead{zhaosm@njupt.edu.cn}
\vspace{10pt}
\begin{indented}
\item[]August 2017
\end{indented}

\begin{abstract}
We propose a multiple pulses phase-matching quantum key distribution  protocol (MPPM-QKD) to exceed the linear key rate bound and to achieve higher error tolerance.
In our protocol, Alice and Bob generate at first their own train pulses (each train should contain $L$ pulses) as well as random
bit sequences, and also encode each pulse of their trains with a randomized phase and a modulation phase.
As the next step, both encoded trains are simultaneously sent to Charlie, who performs an interference detection and may be also an eavesdropper.
After a successful detection is announced by Charlie, Alice and Bob open the randomized phase of each pulse and  keep only communications when the summation of the difference randomized phases at two success detection's time-stamps for Alice and Bob are equal to 0 or $\pi$. Thereafter, Alice and Bob compute the sifted key with the time-stamps.

The above procedure is repeated until both Alice and Bob achieve sufficiently long sifted keys. We can also show that the secret key rate of the proposed QKD protocol can beat the rate-loss limit of so far known QKD protocols when the transmission distance is greater than 250 km.
Moreover,  the proposed protocol has a higher error tolerance, approximately $24\%$, when the transmission distance is 50 km and L = 128.  The secret key rate and the transmission distance of our protocol are superior to that of the round-robin differential-phase-shift quantum key distribution protocol \cite{SASA14}, and
also of the measurement-device-independent quantum key distribution protocol \cite{LO12}, and the secret key rate performance is better in both cases than that of phase-matching quantum key distribution when bit train length is greater than 32.
\end{abstract}

%
\vspace{2pc}

\section{Introduction}

Quantum key distribution (QKD) protocols  allow two distant parties (Alice and Bob) to produce ad share a secret classical key even at the existence of an eavesdropper (Eve) \cite{BB84,GISIN02}.
In both theory and in experiments,
QKD protocols are  expected to be more and more extensively applied in various
practical situations \cite{WANG05,LO12,BRAU12,SASA14,LUCA18,INAM07,ZHAO13,WANGL15,WANGS15,WANG2017,MA17,GAO19}.
However, transmission losses of photons have become one of the major obstacles in practical implementations of QKD protocol \cite{ZHOU16,YIN16}.

Since the first QKD protocol (BB84) was proposed by Bennett \emph{et al.} in 1984,
a variety of other  QKD protocols have successively been presented. Unfortunately,
a larger application impact of these protocols was questioned because it was believed that the key rate ($R$)
of these protocols was bounded with respect to various the transmittance parameters, \cite{CUR04,TAKE14,PIRA17}, $R\leq O(\eta)$,
where $\eta$ is defined as the probability that a photon can be
successfully transmitted through the channel and being detected at the end of the channel. Since
no secure enough key could be distributed when transmitted photons (carrying a quantum information in QKD protocols) are lost in the
channel, the transmittance $\eta$ becomes a natural upper bound of the
secret key rate without trusted relay nodes \cite{SANG11}.

Remarkably, Lucamarini \emph{et al.} have recently proposed a novel phase-encoding QKD
protocol, called as the twin-field QKD (TF-QKD) \cite{LUCA18}, to overcome current rate-distance limits  of QKD  protocols,
without quantum repeaters, and its security proof was presented in Ref.\cite{TAMA18}.
Inspired by the original TF-QKD and its variants, a phase-matching quantum key distribution (PM-QKD) protocol
and also a 'sending or not sending TF-QKD' protocol, were subsequently proposed to overcome  linear
key rate bound with relative phase encoding \cite{MA18} and to tolerate
a large misalignment error \cite{WANG18}, respectively.  Later, a simplified TF-QKD protocol without the post-selection
was introduced and analyzed in \cite{CUI18}. By using vacuum and one-photon state as a qubit,
the TF-QKD could be seen as a measurement-device-independent
QKD (MDI-QKD) \cite{LO12} with a single-photon Bell state measurement \cite{YIN18}.
The physics behind TF-QKD is that Alice and Bob prepare photon-number
superposition remotely via coherent states and post-selection.

In the PM-QKD protocol, Alice (Bob) prepare their weak coherent states $ |\pm \sqrt{\mu}>$ randomly
and add a random phase $\phi(A)(\phi(B))$ to each of
their weak coherent states. Afterwards, they sent states to an untrusted party (Charlie) located somewhere in the
channel. Depending on the measurement performed by Charlie,
Alice and Bob are able to generate the raw key  after a post-selection of the cases satisfying $\phi(A)\approx \phi(B)$.
After a sifting, parameter estimation and key distillation are necessary to be used to generate a final private and secure key.

The signal disturbance caused by Eve's intervention should  be monitored to have a bound on
the potential information leakage. If an
estimation of key parameters of the protocol is required with high precision, the
portion of the signal that had to be sacrificed increases, and that decreases the
efficiency of the protocol. However,  T. Sasaki \emph{et al.} in 2014
proposed in a round robin differential phase shift quantum key distribution protocol \cite{SASA14}, named RRDPS-QKD,
how to obtain the secret key without the requirement to do
high precision estimation of many parameters.
Since the proportion of information an eavesdropper
could have gained about the key can be known in
advance, based on the pre-set system parameters
only, RRDPS-QKD not only simplifies the key
generation process but, more importantly,
can lead to better noise tolerance \cite{WANGS15,WANG2017,CAI09}.

In this paper, we propose a novel PM-QKD protocol without monitoring the disturbance, named multiple pulses phase-matching quantum key distribution protocol(MPPM-QKD). In this protocol, one $L$- pulse sequence (named "train") is prepared by Alice (Bob), individually, which includes only one photon inside. Then, Alice (Bob) randomizes each phase of one pulse in his (her) train and further encodes the phase with ${0,\pi}$ according to the two independent random $L$-bit sequences prepared before.  After that, Alice and Bob send encoded trains to Charlie (an eavesdropper), who is expected to perform the interference detection. Charlie obtains successfully detection when
the clicks happen in exactly two time-stamps and announce time-stamps indices $(i, j)$ of two clicks to Alice and Bob.
Afterward, Alice and Bob match phases of pulses by a post-selection, and compute their sifted key by operations on the
their own $L$-bit sequences with the indices $(i, j)$. After achieving sufficiently large sifted keys,
Alice and Bob obtain a secret key through reconciliation and privacy amplification procedures.

The organization of the paper is as follows. In Section 2, the MPPM-QKD is presented. In Section 3, the secret key rate performance of the proposed MPPM-QKD is analyzed. Finally, Section 4 concludes the paper.

\section{Multiple pulses phase-matching quantum key distribution protocol}

\begin{figure}[!htbp]
  \begin{center}
    \includegraphics[width=1.0\textwidth]{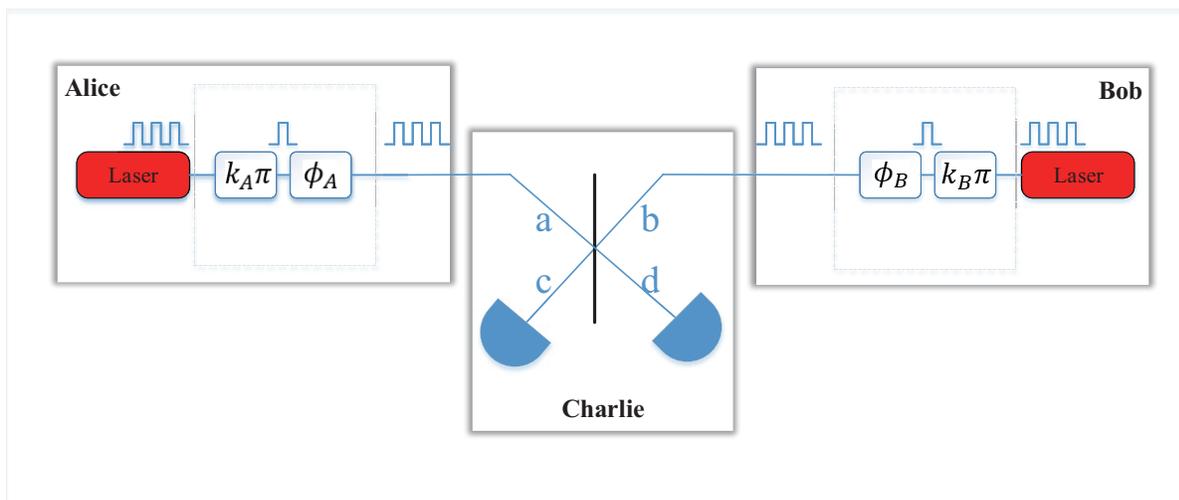}
    \caption{The schematic diagram of the proposed MPPM-QKD protocol.}
  \end{center}
\end{figure}

In the section, we present our new MPPM-QKD protocol.

Fig.1 shows the schematic diagram of the proposed MPPM-QKD protocol. In MPPM-QKD protocol,  two authorized users, Alice and Bob,
generate two trains pulses (each train contains $L$ pulses) and two $L$-bit random
bit-sequences $k_A=k_A^1k_A^2\cdots k_A^L \in \{0,1\}^{\otimes L}$ and $k_B=k_B^1k_B^2\cdots k_B^L \in \{0,1\}^{\otimes L}$.
Here, and in the following, the lower-case $A$ denotes Alice and the lower-case $B$ represents Bob.
They then perform a randomized phase coding $\phi_{A(B)}\in[0,2\pi)$
and phase modulation on every pulse of their trains, individually.
The phase modulation rule is as follows: the encoding of the pulse phase to $0$ when $k_{A(B)}=0$ and an encoding of the pulse phase to $\pi$ when $k_{A(B)}=1$.
After that, both Alice and Bob send simultaneously their encoded trains to Charlie,
who may be an untrusted Eve, and is expected to perform the interference detection at his side.
After the interference detection has been carried out, Charlie announces the success detection time-stamps $(m,n), 0 \leq m,n \leq (L-1)$.
Observe that only two clicks happened when the interference detection is defined as a success detection.
After the success detection, Alice and Bob  open their randomized phase coding for each pulse.
Only when $\left |\phi_{A}^{m}+\phi_{B}^{n}-\phi_{A}^{n}-\phi_{B}^{m}\right |=0$ or $\pi$, Alice and Bob keep on communication,
and compute the sifted key with the success detection time-stamps.
The procedures is repeated until Alice and Bob obtain enough sifted keys.

\noindent The proposed protocol can be described in details as follows.
\begin{enumerate}
\item Alice prepares her $L$-pulse train. (Here and in the following, "$L$ successive pulses" will be used as the name of the train.) Alice then generates a $L$-bit random
sequence $k_A=k_A^1k_A^2\cdots k_A^L \in \{0,1\}^{\otimes L}$.  For the $m$-th pulse of the train ($0\leq m \leq (L-1)$),
Alice encodes a random phase $\phi_A^m\in[0,2\pi)$ on the pulse,
and applies the phase modulation $0$ (or $\pi$) on the pulse according to the value of $k_A^m$. Similarly, Bob prepares and encodes his $L$-pulse train the same way as Alice does. For his $n$-th pulse, Bob randomly generates a random phase $\phi_B^n \in[0,2\pi)$, and applies phase modulation $0$ ($\pi$) according to $k_B^n$ value, where $k_B=k_B^1k_B^2\cdots k_B^L \in \{0,1\}^{\otimes L}$. Let us now consider a simple case where both Alice and Bob have exactly one photon in their $L$-pulse trains. The states of the two photons can now be expressed by
\begin{eqnarray} \label{eqn1}
 \frac{1}{\sqrt{L}}\sum_{m=1}^{L}e^{i(\phi_A^m+k_A^{m}\pi)}\hat{a}_{m}^{+}\mid0>, \\ \nonumber
 \frac{1}{\sqrt{L}}\sum_{n=1}^{L}e^{i(\phi_B^n+k_B^{n}\pi)}\hat{b}_{n}^{+}\mid0>.
\end{eqnarray}
where $i$ is an image unit $i=\sqrt{(-1)}$. $\hat{a}_{m}^{+}, \hat{b}_{n}^{+}$ are the creation operators on Alice's $m$-th and Bob's $n$-th pulse at a's and b's paths in Fig.1, respectively.
\item Alice and Bob send their encoded trains to Charlie, who is expected to perform an interference measurement and record  detector's clicks. Since there are two photons in a block, one from Alice and one from Bob, Charlie  obtains at most two detection clicks. He then post-selects the block where there are exactly two clicks. A success detection is now defined as wo clicks in the interference measurement. If there is only one click, the detection result is discarded in the proposed  protocol.
    After the interference and Charlie's post-selection, the quantum state at the two detectors become one of the four states,
\begin{eqnarray} \label{eqn2}
 (1+ \cos(\phi_{A}^{m} +k_A^{m}\pi +\phi_{B}^{n}+k_B^{n}\pi \\ \nonumber
   - \phi_{A}^{n}-k_A^{n}\pi -\phi_{B}^{m}-k_B^{m}\pi ))c_{m}^{+}c_{n}^{+}\mid0>, \\ \nonumber
 (1-\cos(\phi_{A}^{m} +k_A^{m}\pi +\phi_{B}^{n}+k_B^{n}\pi  \\ \nonumber
 -\phi_{A}^{n}-k_A^{n}\pi -\phi_{B}^{m}-k_B^{m}\pi ))c_{m}^{+}d_{n}^{+}\mid0>,    \\ \nonumber
 (1-\cos(\phi_{A}^{m} +k_A^{m}\pi +\phi_{B}^{n}+k_B^{n}\pi  \\ \nonumber
 -\phi_{A}^{n}-k_A^{n}\pi -\phi_{B}^{m}-k_B^{m}\pi ))d_{m}^{+}c_{n}^{+}\mid0>,    \\ \nonumber
 (1+\cos(\phi_{A}^{m} +k_A^{m}\pi +\phi_{B}^{n}+k_B^{n}\pi  \\ \nonumber
 -\phi_{A}^{n}-k_A^{n}\pi -\phi_{B}^{m}-k_B^{m}\pi ))d_{m}^{+}d_{n}^{+}\mid0>.
\end{eqnarray}
The detail is deduced in Appendix A. Here, $\hat{c}_{m}^{+}$ and $\hat{d}_{m}^{+}$ mean the creation operators at $m$ time-bin at c and d paths to the two detectors, respectively.

\item Charlie announces time-stamps $(m,n)$ of  two clicks when the interference measurement detection is a success. After Charlie's announcement, Alice and Bob open their random phases of the corresponding pulses $\phi_{A}^{m},\phi_{A}^{n},\phi_{B}^{m},\phi_{B}^{n}$.

\item Only if $\left |\phi_{A}^{m}+\phi_{B}^{n}-\phi_{A}^{n}-\phi_{B}^{m}\right |=0$ or $\pi$, will Alice and Bob keep up communications. Since $k_A^{m},k_B^{n},k_A^{n},k_B^{m}$ is $\in \{0,1\}$, the value of $\left |k_A^{m}+k_B^{n}-k_A^{n}-k_B^{m}\right |$ should be $0,1 $ or $2$. That is, $\left |k_A^{m}+k_B^{n}-k_A^{n}-k_B^{m}\right |=1$ or $\left |k_A^{m}+k_B^{n}-k_A^{n}-k_B^{m}\right |\neq 1$.

    Let us Consider now cases that $\left |\phi_{A}^{m}+\phi_{A}^{n}-\phi_{B}^{m}-\phi_{B}^{n}\right |=0$.
    If $\left |k_A^{m}+k_B^{n}-k_A^{n}-k_B^{m}\right |=0$ or $2$,  two considered clicks should be triggered by the same detector;
    Of course,  clicks should be triggered by different detectors when $\left |k_A^{m}+k_B^{n}-k_A^{n}-k_B^{m}\right |=1$.
    If two clicks are triggered by the same detector, $k_A^{m}\oplus k_A^{n}$ then should be equal to $k_B^{m}\oplus k_B^{n}$.
    Therefore, Alice and Bob would have their sifted key as $s_{A}=k_A^{m}\oplus k_A^{n}$, $s_{B}=k_B^{m}\oplus k_B^{n}$.
    On the other hand, when both clicks are triggered by different detectors, $!(k_A^{m}\oplus k_A^{n})$ is  equal to $k_B^{m}\oplus k_B^{n}$,
    Alice and Bob could get in such a case their sifted key as $s_{A}=k_A^{m}\oplus k_A^{n}$, $s_{B}=!(k_B^{m}\oplus k_B^{n})$.

    Conversely, when $\left |\phi_{A}^{m}+\phi_{B}^{n}-\phi_{A}^{n}-\phi_{B}^{m}\right |=\pi$,  clicks are triggered by the same detector when $!(k_A^{m}\oplus k_A^{n})=k_B^{m}\oplus k_B^{n}$. Alice and Bob would get in such a case sifted keys as $s_{A}=k_A^{m}\oplus k_A^{n}$, $s_{B}=!(k_B^{m}\oplus k_B^{n})$; Two clicks were triggered by different detectors when $k_A^{m}\oplus k_A^{n}=k_B^{m}\oplus k_B^{n}$, Alice and Bob could compute in such a case their sifted keys as $s_{A}=k_A^{m}\oplus k_A^{n}$, $s_{B}=k_B^{m}\oplus k_B^{n}$.
    (See the Appendix B for details.) This way, Alice and Bob would obtain a sifted key without estimating  signal disturbance in quantum channels.

\item Alice and Bob repeat steps (i)-(iv) several times until they have enough
of sifted keys. Afterwards, they perform an error correction and privacy amplification on the sifted key bits to produce a fully secret key.
\end{enumerate}
In the single photon case, it can be shown that Charlie,  seen as an eavesdropper, cannot
distinguish whether the photon causing a click belongs to Alice or Bob when he post-selects the block where the two clicks happen at $m$ and $n$ time-bins.

Suppose that Alice's photon is at $m$ time-bin, Bob's photon is at $n$ time-bin. Alice and Bob encode their key information ($k_{A}^{m}, k_{B}^{n}$) and randomize the pulse phases ($\phi_{A}^{m}, \phi_{B}^{n}$) on the $m$-th pulse and $n$-th pulse of their trains, respectively.
Afterwards, they send the encoded quantum states to Charlie who is expected to
apply to it an interference measurement and to announce whether $\left |k_A^{m}+k_B^{n}-k_A^{n}-k_B^{m}\right |=1$ or $\left |k_A^{m}+k_B^{n}-k_A^{n}-k_B^{m}\right |\neq1$. Observe that there are sixteen possible output states that could be sent to Charlie.
We suppose now that $\left |\phi_{A}^{m}+\phi_{B}^{n}-\phi_{A}^{n}-\phi_{B}^{m}\right |=0$ and that clicks were triggered by different detectors,
so that Charlie will get $\left |k_A^{m}+k_B^{n}-k_A^{n}-k_B^{m}\right |=1$.
Unfortunately, Charlie can  get only one of the eight possible output states from the result $\left |k_A^{m}+k_B^{n}-k_A^{n}-k_B^{m}\right |=1$. Here we describe them as $\mid\Psi _{0,0,0,\pi}>$, $\mid\Psi _{0,0,\pi,0}>$, $\mid\Psi _{0,\pi,0,0}>$, $\mid\Psi _{0,\pi,\pi,\pi}>$, $\mid\Psi _{\pi,0,0,0}>$, $\mid\Psi _{\pi,0,\pi,\pi}>$, $\mid\Psi _{\pi,\pi,0,\pi}>$, $\mid\Psi _{\pi,\pi,\pi,0}>$. From these results, Charlie cannot determine,
in principle,  whether phases encoded by Alice on her $m$-th pulse and $n$-th pulse (or  phases encoded by Bob on his $m$-th pulse and $n$-th pulse) are 0 or $\pi$.
Thus, Charlie also can not get the final key as $k_{E}=!(k_A^{m}\oplus k_A^{n})=k_B^{m}\oplus k_B^{n}$.


In practice, a single-photon state source is often replaced by a weak laser pulse, which can be described by a coherent state. That works as follows.
Alice generates a coherent state pulse, encodes her information and randomizes each pulse of the train.
The output quantum state is now $\frac{1}{\sqrt{L}}\sum_{m=1}^{L}\{e^{i(\phi_A^m+k_A^{m}\pi)}\sum_{k=0}^{+\infty}\sqrt{e^{-\mu}\frac{\mu^{k}}{k!}}\mid k>\}$, where $k$ is the photon number and $\mu$ is the mean photon number or the light intensity.
Bob does the same state preparation as Alice.  His quantum output state is $\frac{1}{\sqrt{L}}\sum_{n=1}^{L}\{e^{i(\phi_B^n+k_B^{n}\pi)}\sum_{k=0}^{+\infty}\sqrt{e^{-\mu}\frac{\mu^{k}}{k!}}\mid k>\}$.
When the phase is randomized, it can be shown that the state of the whole train can be described by a statistical mixture of Fock states, whose photon numbers follows a Poisson distribution. Similarly to the single-photon case, Alice's key information is encoded into the relative phases between  two pulses. Of course, it is possible to have multi-photons components in Alice and Bob's individual pulse trains, that inevitably effects Charlie's post processing strategy. If Charlie gets two or more detection clicks in a block, he could randomly choose the two time stamps $m$ and $n$ and announce the time-bins $(m,n)$ pairs to Alice and Bob. Otherwise, he discards the result. In this way, Alice and Bob can figure out the phase relationship between $m$ and $n$ as in the single photon case.

\section{Secret key rate}
In the section, we discuss the secret key rate of the proposed protocol.

We will also use the Gottesman-Lo-Lutkenhaus-Preskill (GLLP) formula \cite{GO04} for the secret key rate derivations.
It will be shown that the secret key rate is
\begin{equation} \label{eqn3}
 R=\frac{Q_u}{L}(1-H_{PA}-H_{EC}).
\end{equation}
where $L$ stands for the length of pulses and $H_{EC}=f\cdot H(E_u)$ accounts for the cost of an error correction and $H_{PA}=H(e_p)$ stands for privacy amplification. In addition, $E_u$ denotes the bit error rate and $e_p$  the phase error rate, $f$ is the error correction efficiency, and $H(x)=-x\log_{2}x-(1-x)\log_{2}(1-x)$ is the binary Shannon entropy function.

For the weak coherent sources, the probability of photon numbers $n$ for the quantum state follows a Poisson distribution, that is,
\begin{equation} \label{eqn4}
P_n=e^{-\mu}\frac{\mu^{n}}{n!}.
\end{equation}
where $n$ is the photon number and $\mu$ is the mean photon number or the light intensity. In the simulation, the yield $Y_n$ and error rates $e_n$ of the n-photon component are given by \cite{MA18},
\begin{eqnarray} \label{eqn5}
Y_n&=1-(1-2Y_{0})(1-\eta)^{n},    \\
e_n&=\frac{Y_{0}(1-\eta)^{n}+e_{d}[1-(1-\eta)^{n}]}{Y_n}.
\end{eqnarray}
where $Y_{0}$ is the dark-count rate, $\eta$ represents the overall transmittance of the system and $e_{d}$ is the misalignment probability.

For the weak coherent source whose mean photon number or the light intensity is $\mu$,  pulses total gain is $Q_u$ and the quantum bit error rate(QBER) $E_u$ can be expressed as
\begin{eqnarray} \label{eqn6}
 Q_u=\sum_{n=0}^{\infty} P_nY_n   \nonumber \\
    =\sum_{n=0}^{\infty} e^{-\mu}\frac{\mu^{n}}{n!}[1-(1-2p_{d})(1-\eta)^{n}]   \nonumber \\
    =1-(1-2p_{d})e^{-\mu\eta},    \\
 E_u=\sum_{n=0}^{\infty} P_ne_nY_n    \nonumber \\
    =\sum_{n=0}^{\infty} 1-(1-2p_{d})(1-\eta)^{n}[p_{d}(1-\eta)^{n}+e_{d}[1-(1-\eta)^{n}]]  \nonumber \\
    =e_{d}+(p_{d}-e_{d})e^{-\mu\eta}.
\end{eqnarray}

The detection click is classified as being of two types: a success detection and a failing detection. Two clicks from the same detector at different times, and two clicks from  two detectors at different times, are considered as success detections.
For a success detection, the phase error rate can be calculated using the RRDPS-QKD protocol. We will
set $v_{th}$ as a threshold photon number for the source. The phase error rate will be bounded by $1/2$ when the mean photon number is greater than the threshold. Since the phase error
rate increases with the photon number, one can consider the worst case scenario to be the case where the
losses are all contributed from the lowest photon numbers.
For the quantum signal containing the mean photon  number less than $v_{th}$  the phase error rate will be bounded \cite{Guan15} by
\begin{equation} \label{eqn7}
e_p=\frac{e\_src}{Q_u}+(1-\frac{e\_src}{Q_u})*\frac{1-(1-\frac{2}{(L-1)})^{v_{th}}}{4}+\frac{\mu\eta}{2(1-\mu\eta)}.
\end{equation}
where $e\_src=P(n> v_{th})$ is the probability that the  photon number is greater than the threshold. $Q_u$ will be the gain of the experiment, $\eta$ represents the overall transmittance of the system, $\mu$ will be the mean photon number.
The phase error rate will be calculated corresponding to the probability of more than $v_{th}$ photons: The probability of having less than $v_{th}$ photons and the probability that two or more photons simultaneously enter the same detector at the same time.
Therefore, the final key generation formula of the MPPM-QKD protocol is
\begin{equation} \label{eqn8}
R=\frac{1}{L}[Q_u-Q_u*f*h(E_u)-Q_u*h(e_p)].
\end{equation}

\section{Simulation and Analysis}
In this section, we deal with a numerical simulation of the secret key rate (SKR),  and compare the SKR performance of the proposed protocol with tat of MDI-QKD and RRDPS-QKD protocols.
\begin{table}[htbp]
\centering
\caption{Values of key parameters of our numerical simulation.}
\begin{indented}
\lineup
\item[]\begin{tabular}{@{}*{5}{l}}
\br
$\0\0Y_0$& $\eta_B$& $e_d$& \m$f$& \m$\alpha/(dB/km)$\cr
\mr
\0\0$10^{-9}\times L$& $19\%$& $1.5\%$& \0$1.16$& \0\0$0.2$ \cr
\br
\end{tabular}
\end{indented}
\label{TABLE_I}
\end{table}

Key parameters of our numerical simulation are given in Table~\ref{TABLE_I}.
Channels between Alice and Charlie, as well as between Bob and Charlie, are supposed to be identical.

\begin{figure}[!htbp]
  \begin{center}
    \includegraphics[width=1.0\textwidth]{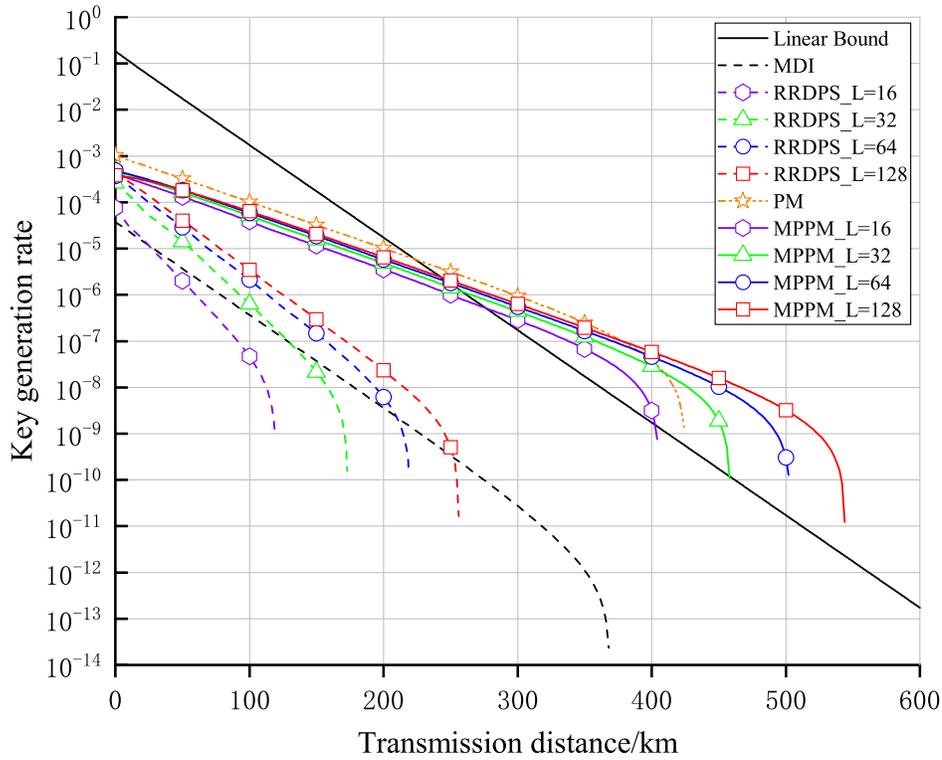}
    \caption{Secret key rate of the proposed MPPM-QKD protocol for different
transmission distances $L$, compared with those for PM-QKD, MDI-QKD, RRDPS-QKD protocols.}
  \end{center}
\end{figure}
Fig. 2 shows secret key rates of the proposed MPPM-QKD protocol with respect different transmission distances $L$, and comparing them  with those of PM-QKD, MDI-QKD, RRDPS-QKD protocols. From the figure one can see that the proposed MPPM-QKD protocol is able to exceed
the linear key rate bound when the transmission distance is greater than $250$ km. The longer the transmission distance is, the smaller the secret key rate is. For the same transmission distance, the longer train is, the higher the secret secret key rate is.

When compared with the PM-QKD protocol, MPPM-QKD protocol can be used for a longer transmission distance when $L$ is greater than 32.
In comparison with the MDI-QKD protocol, MPPM-QKD protocol can function well at longer transmission distance and a higher secret key rate.
In addition, MPPM-QKD protocol has a better performance in comparison with RRDPS-QKD protocol for the same time length $L$.
For the key rate $R=10^{-8}$ and $L=16$,  the longest practical transmission distance of the MPPM-QKD protocol reliability is approaching $400$ km, whereas MDI-QKD protocol has it lower than $200$ km and the RRDPS-QKD is upper bounded by $100$ km. Moreover, the key rate is increased by $2\sim 4$ orders of magnitude when the transmission distance is $200$ km,


\begin{figure}[!htbp]
  \centering
  \subfigure[MPPM-QKD and PM-QKD]{\includegraphics[width=3in]{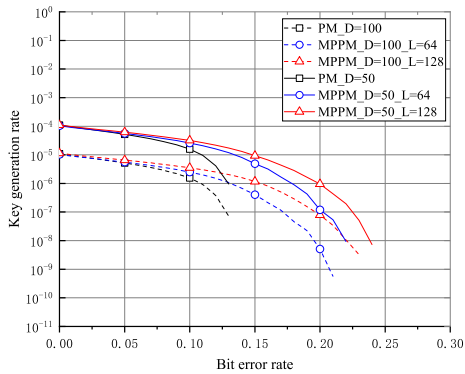}}
  \subfigure[MPPM-QKD and RRDPS-QKD]{\includegraphics[width=3in]{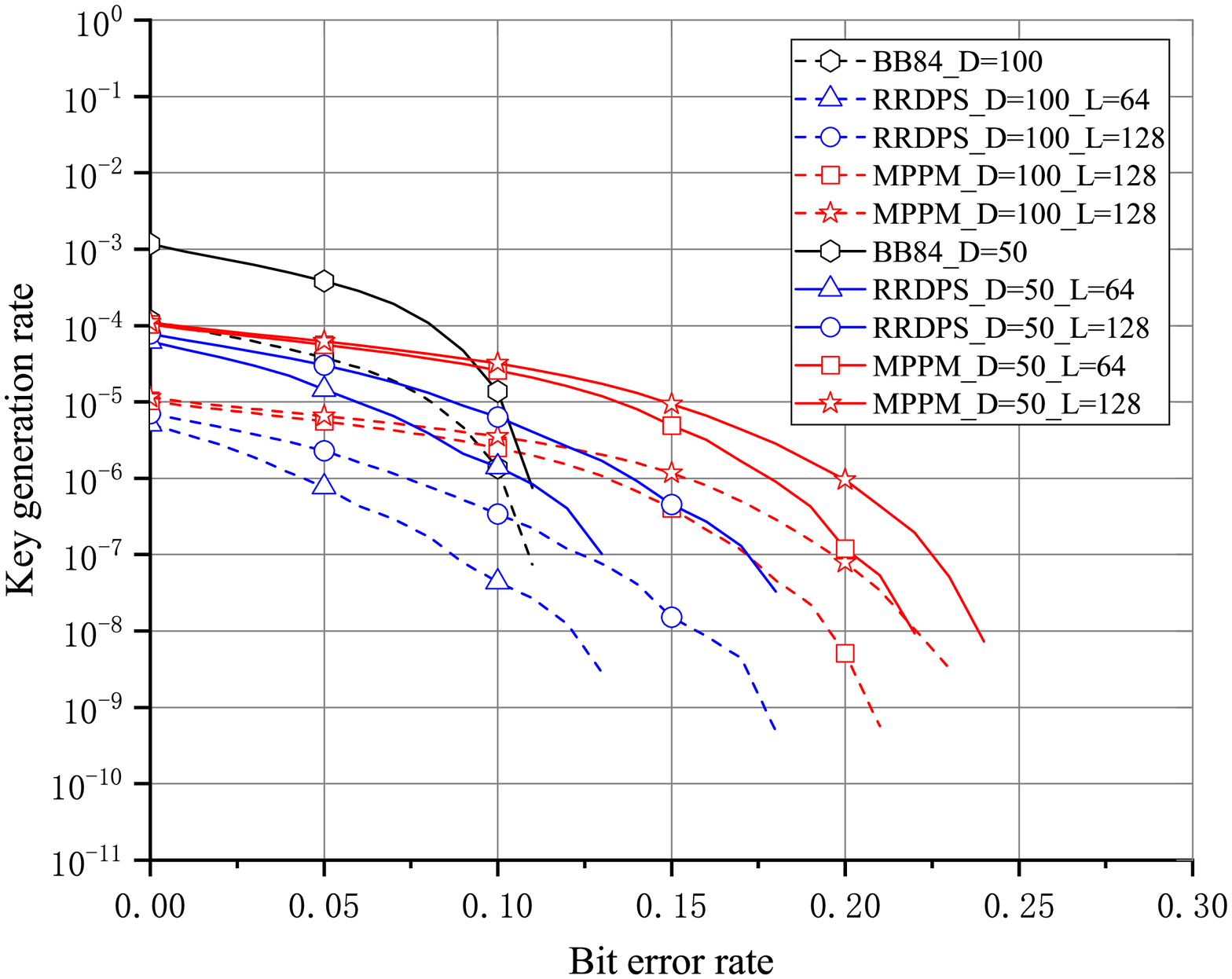}}
  \caption{Key generatin rates of the MPPM-QKD, PM-QKD and RRDPS-QKD protocols versus bit error rates.}
\end{figure}
Fig.3 shows the secret secret key rate of the MPPM-QKD protocol versus the bit error rate, together with those performances of PM-QKD and RRDPS-QKD protocols, when the transmission distances were setup to $50km$ and $100km$. Fig.3(a) shows the comparison with the PM-QKD protocol, while Fig.3(b) shows the comparison with the RRDPS-QKD protocol.
From the figure, we can see that the secret secret key rate decrease with the increase of the bit error rate. When $L=128$, the performance of MPPM-QKD protocol was better than that of PM-QKD. When compared with the RRDPS-QKD protocol, MPPM-QKD protocol has a better secret key rate with the same bit error rate, same transmission distance, and the same train length.
When the transmission distance is 50 km and $L=128$, the highest bit error rate of the MPPM-QKD protocol is approximately $24\%$, which is higher than $18\%$ that is the highest bit error rate of the RRDPS-QKD protocol.

\section{Conclusions}
In this paper, we have proposed a new MPPM-QKD protocol, which has the advantages comparing with  MP-QKD and RRDPS-QKD protocols. The proposed QKD protocol can exceed the linear key rate bound in general QKD protocols like MP-QKD, and does not monitor signal disturbance in the used quantum channel.  The numerical results have demonstrated that our MPPM-QKD protocol not only have exceeded the well-known rate-loss limit of known QKD protocols, but have also a better performance on secrecy of the secret key rate and also concerning transmission distance at which it can be used.  The proposed MPPM-QKD protocol has a better secret key generation performance also in comparison with the RRDPS-QKD protocol for the same train length $L$, and have achieved a longer usable transmission distance when the length of train $L$ is greater than 32 in comparison with the MP-QKD protocol. In comparison with RRDPS-QKD, MPPM-QKD protocols it has also a higher error tolerance.
The transmission distance of the proposed MPPM-QKD protocol approaches  450 km when the secret key rate $R$ has been $10^{-8}$, and the usable length $L$ has been $128$.
The secret key rate has increased by two-three orders of magnitude, compared with those of the RRDPS-QKD protocol and MDI-QKD protocol, when the transmission distance has been set up to $200km$. Moreover, the highest bit error rate has been approximate  $25\%$ when the transmission distance has been 50 km and $L=128$.

\section *{Acknowledgements}
The paper is supported in part by the National Natural Science Foundation of China (61271238 and 61475075).

\section *{Reference}

\textcolor[rgb]{1.00,0.00,0.00}{\section *{Appendix A}}
\renewcommand{\theequation}{A\arabic{equation}}
The quantum states of Alice and Bob are given by
\begin{equation}\label{eqn9}
 \frac{1}{\sqrt{L}}\sum_{m=1}^{L}e^{i(\phi_A^m+k_A^{m}\pi)}\hat{a}_{m}^{+}\mid0>,
 \frac{1}{\sqrt{L}}\sum_{n=1}^{L}e^{i(\phi_B^n+k_B^{n}\pi)}\hat{b}_{n}^{+}\mid0>.
\end{equation}
respectively, where $\hat{a}_{m}^{+}, \hat{b}_{m}^{+}$ are the creation operator of Alice's and Bob's m-th position at a and b paths, respectively, as shown in Fig.1. After the interference by the beam splitter which replaces $\hat{a}_{m}^{+}$ by $(\hat{c}_{m}^{+}+\hat{d}_{m}^{+})/\sqrt{2}$ and $\hat{b}_{n}^{+}$ by $(\hat{c}_{m}^{+}-\hat{d}_{n}^{+})/\sqrt{2}$, the output result becomes
\begin{eqnarray} \label{labe10}
(\frac{1}{\sqrt{L}}\sum_{m=1}^{L}e^{i(\phi_A^m+k_A^{m}\pi)}\hat{a}_{m}^{+})(\frac{1}{\sqrt{L}} \nonumber
\sum_{n=1}^{L}e^{i(\phi_B^n+k_B^{n}\pi)}\hat{b}_{n}^{+})    \\ \nonumber
=(\frac{1}{\sqrt{2L}}\sum_{m=1}^{L}e^{i(\phi_A^m+k_A^{m}\pi)}(\hat{c}_{m}^{+}+\hat{d}_{m}^{+}))(\frac{1}{\sqrt{2L}}\sum_{n=1}^{L}e^{i(\phi_B^n
+k_B^{n}\pi)}(\hat{c}_{n}^{+}-\hat{d}_{n}^{+}))    \\   \nonumber
=(\frac{1}{\sqrt{2L}}\sum_{m=1}^{L}e^{i(\phi_A^m+k_A^{m}\pi)}\hat{c}_{m}^{+}+\frac{1}{\sqrt{2L}}\sum_{m=1}^{L}e^{i(\phi_A^m+k_A^{m}\pi)} \nonumber
\hat{d}_{m}^{+})(\frac{1}{\sqrt{2L}}\sum_{n=1}^{L}e^{i(\phi_B^n+k_B^{n}\pi)}\hat{c}_{n}^{+} \nonumber   \\
-\frac{1}{\sqrt{2L}}\sum_{n=1}^{L}e^{i(\phi_B^n+k_B^{n}\pi)}\hat{d}_{n}^{+}) \\ \nonumber
=e^{i(\phi_A^1+k_A^{1}\pi+\phi_B^{1}+k_B^{1}\pi)}\hat{c}_{1}^{+}\hat{c}_{1}^{+}+...+e^{i(\phi_A^1+k_A^{1}\pi+\phi_B^{L}+k_B^{L}\pi)}\hat{c}_{1}^{+}\hat{c}_{L}^{+}
\nonumber   \\
-e^{i(\phi_A^1+k_A^{1}\pi+\phi_B^{1}+k_B^{1}\pi)}\hat{c}_{1}^{+}\hat{d}_{1}^{+}-...-e^{i(\phi_A^1+k_A^{1}\pi+\phi_B^{L}+k_B^{L}\pi)}\hat{c}_{1}^{+}\hat{d}_{L}^{+}
\nonumber   \\
+e^{i(\phi_A^2+k_A^{2}\pi+\phi_B^{1}+k_B^{1}\pi)}\hat{c}_{2}^{+}\hat{c}_{1}^{+}+...+e^{i(\phi_A^2+k_A^{2}\pi+\phi_B^{L}+k_B^{L}\pi)}\hat{c}_{2}^{+}\hat{c}_{L}^{+}
\nonumber   \\
-e^{i(\phi_A^2+k_A^{2}\pi+\phi_B^{1}+k_B^{1}\pi)}\hat{c}_{2}^{+}\hat{d}_{1}^{+}-...-e^{i(\phi_A^2+k_A^{2}\pi+\phi_B^{L}+k_B^{L}\pi)}\hat{c}_{2}^{+}\hat{d}_{L}^{+}
\nonumber   \\
+...
\nonumber   \\
+e^{i(\phi_A^L+k_A^{L}\pi+\phi_B^{1}+k_B^{1}\pi)}\hat{c}_{L}^{+}\hat{c}_{1}^{+}+...+e^{i(\phi_A^L+k_A^{L}\pi+\phi_B^{L}+k_B^{L}\pi)}\hat{c}_{L}^{+}\hat{c}_{L}^{+}
\nonumber   \\
-e^{i(\phi_A^L+k_A^{L}\pi+\phi_B^{1}+k_B^{1}\pi)}\hat{c}_{L}^{+}\hat{d}_{1}^{+}-...-e^{i(\phi_A^L+k_A^{L}\pi+\phi_B^{L}+k_B^{L}\pi)}\hat{c}_{L}^{+}\hat{d}_{L}^{+}
\nonumber   \\
+e^{i(\phi_A^1+k_A^{1}\pi+\phi_B^{1}+k_B^{1}\pi)}\hat{d}_{1}^{+}\hat{c}_{1}^{+}+...+e^{i(\phi_A^1+k_A^{1}\pi+\phi_B^{L}+k_B^{L}\pi)}\hat{d}_{1}^{+}\hat{c}_{L}^{+}
\nonumber   \\
-e^{i(\phi_A^1+k_A^{1}\pi+\phi_B^{1}+k_B^{1}\pi)}\hat{d}_{1}^{+}\hat{d}_{1}^{+}-...-e^{i(\phi_A^1+k_A^{1}\pi+\phi_B^{L}+k_B^{L}\pi)}\hat{d}_{1}^{+}\hat{d}_{L}^{+}
\nonumber   \\
+e^{i(\phi_A^2+k_A^{2}\pi+\phi_B^{1}+k_B^{1}\pi)}\hat{d}_{2}^{+}\hat{c}_{1}^{+}+...+e^{i(\phi_A^2+k_A^{2}\pi+\phi_B^{L}+k_B^{L}\pi)}\hat{d}_{2}^{+}\hat{c}_{L}^{+}
\nonumber   \\
-e^{i(\phi_A^2+k_A^{2}\pi+\phi_B^{1}+k_B^{1}\pi)}\hat{d}_{2}^{+}\hat{d}_{1}^{+}-...-e^{i(\phi_A^2+k_A^{2}\pi+\phi_B^{L}+k_B^{L}\pi)}\hat{d}_{2}^{+}\hat{d}_{L}^{+}
\nonumber   \\
+...
\nonumber   \\
+e^{i(\phi_A^L+k_A^{L}\pi+\phi_B^{1}+k_B^{1}\pi)}\hat{d}_{L}^{+}\hat{c}_{1}^{+}+...+e^{i(\phi_A^L+k_A^{L}\pi+\phi_B^{L}+k_B^{L}\pi)}\hat{d}_{L}^{+}\hat{c}_{L}^{+}
\nonumber   \\
-e^{i(\phi_A^L+k_A^{L}\pi+\phi_B^{1}+k_B^{1}\pi)}\hat{d}_{L}^{+}\hat{d}_{1}^{+}-...-e^{i(\phi_A^L+k_A^{L}\pi+\phi_B^{L}+k_B^{L}\pi)}\hat{d}_{L}^{+}\hat{d}_{L}^{+} \nonumber   \\
=\frac{1}{2L}\sum_{1 \le m \le L}^{}e^{i(\phi_A^m+k_A^{m}\pi+\phi_B^{m}+k_B^{m}\pi)}\hat{c}_{m}^{+}\hat{c}_{m}^{+}+\frac{1}{2L}\sum_{1 \le m \le  \nonumber L}^{}e^{i(\phi_A^{m}+k_A^{m}\pi+\phi_B^{m}+k_B^{m}\pi)}\hat{d}_{m}^{+}\hat{d}_{m}^{+}  \nonumber \\
+\frac{1}{2L}\sum_{1 \le m < n \le L}^{}(e^{i(\phi_A^{m}+k_A^{m}\pi+\phi_B^{n}+k_B^{n}\pi)}+e^{i(\phi_A^{n}+k_A^{n}\pi+\phi _B^{m}+k_B^{m}\pi)})\hat{c}_{m}^{+}\hat{c}_{n}^{+}  \nonumber \\
-\frac{1}{2L}\sum_{1 \le m < n \le L}^{}(e^{i(\phi_A^{m}+k_A^{m}\pi+\phi_B^{n}+k_B^{n}\pi)}+e^{i(\phi_A^{n}+k_A^{n}\pi+\phi _B^{m}+k_B^{m}\pi)})\hat{d}_{m}^{+}\hat{d}_{n}^{+}  \nonumber \\
+\frac{1}{2L}\sum_{1 \le m < n \le L}^{}(-e^{i(\phi_A^{m}+k_A^{m}\pi+\phi_B^{n}+k_B^{n}\pi)}+e^{i(\phi_A^{n}+k_A^{n}\pi+\phi _B^{m}+k_B^{m}\pi)})\hat{c}_{m}^{+}\hat{d}_{n}^{+}  \nonumber \\
+\frac{1}{2L}\sum_{1 \le m < n \le L}^{}(e^{i(\phi_A^{m}+k_A^{m}\pi+\phi_B^{n}+k_B^{n}\pi)}-e^{i(\phi_A^{n}+k_A^{n}\pi+\phi _B^{m}+k_B^{m}\pi)})\hat{d}_{m}^{+}\hat{c}_{n}^{+} \nonumber
\end{eqnarray}

There are 6 items in the output result, and only the last four items are kept, since the detection results are discarded when there is only one click in the detectors.
The four items' probabilities are,
\begin{eqnarray} \label{labe11}
\nonumber
P(c_{m}^{+}c_{n}^{+})=<c_{m}^{+}c_{n}^{+}\mid c_{m}^{+}c_{n}^{+}>=1+\cos(\phi_A^{m} +k_A^{m}\pi +\phi_B^{n}+k_B^{n}\pi \\ \nonumber
-\phi_A^{n}-k_A^{n}\pi -\phi_B^{m}-k_B^{m}\pi )  \\ \nonumber
P(c_{m}^{+}d_{n}^{+})=<c_{m}^{+}d_{n}^{+}\mid c_{m}^{+}d_{n}^{+}>=1-\cos(\phi_A^{m} +k_A^{m}\pi +\phi_B^{n}+k_B^{n}\pi \\ \nonumber
-\phi_A^{n}-k_A^{n}\pi -\phi_B^{m}-k_B^{m}\pi )  \\
P(d_{m}^{+}c_{n}^{+})=<d_{m}^{+}c_{n}^{+}\mid d_{m}^{+}c_{n}^{+}>=1-\cos(\phi_A^{m} +k_A^{m}\pi +\phi_B^{n}+k_B^{n}\pi \\ \nonumber
-\phi_A^{n}-k_A^{n}\pi -\phi_B^{m}-k_B^{m}\pi )  \\  \nonumber
P(d_{m}^{+}d_{n}^{+})=<d_{m}^{+}d_{n}^{+}\mid d_{m}^{+}d_{n}^{+}>=1+\cos(\phi_A^{m} +k_A^{m}\pi +\phi_B^{n}+k_B^{n}\pi  \nonumber \\
-\phi_A^{n}-k_A^{n}\pi -\phi_B^{m}-k_B^{m}\pi )  \nonumber
\end{eqnarray}

\section *{Appendix B}
\renewcommand{\thetable}{B\arabic{table}}
\begin{table}[!htbp]
\caption{The results of detectors and the sifted key of Alice and Bob.}
\begin{tabular}{|c|c|c|c|c|c|c|c|}
\hline
\multicolumn{2}{|c|}{Alice Code}&\multicolumn{2}{|c|}{Bob Code}&\multicolumn{4}{|c|}{}\\
\hline
$k_A^{m}$&$k_A^{n}$&$k_B^{m}$&$k_B^{n}$&$\left |k_A^{m}+k_B^{n}-k_A^{n}-k_B^{m}\right |$&Detector results&$s_{A}$&$s_{B}$ \\
\hline
0& 0& 0& 0& 0& SAME& $k_A^{m}\oplus k_A^{n}=0$& $k_B^{m}\oplus k_B^{n}=0$\\
\hline
0& 0& 0& 1& 1& DIFF& $k_A^{m}\oplus k_A^{n}=0$& $!(k_B^{m}\oplus k_B^{n})=0$\\
\hline
0& 0& 1& 0& 1& DIFF& $k_A^{m}\oplus k_A^{n}=0$& $!(k_B^{m}\oplus k_B^{n})=0$\\
\hline
0& 0& 1& 1& 0& SAME& $k_A^{m}\oplus k_A^{n}=0$& $k_B^{m}\oplus k_B^{n}=0$\\
\hline
0& 1& 0& 0& 1& DIFF& $k_A^{m}\oplus k_A^{n}=1$& $!(k_B^{m}\oplus k_B^{n})=1$\\
\hline
0& 1& 0& 1& 0& SAME& $k_A^{m}\oplus k_A^{n}=1$& $k_B^{m}\oplus k_B^{n}=1$\\
\hline
0& 1& 1& 0& 2& SAME& $k_A^{m}\oplus k_A^{n}=1$& $k_B^{m}\oplus k_B^{n}=1$\\
\hline
0& 1& 1& 1& 1& DIFF& $k_A^{m}\oplus k_A^{n}=1$& $!(k_B^{m}\oplus k_B^{n})=1$\\
\hline
1& 0& 0& 0& 1& DIFF& $k_A^{m}\oplus k_A^{n}=1$& $!(k_B^{m}\oplus k_B^{n})=1$\\
\hline
1& 0& 0& 1& 2& SAME& $k_A^{m}\oplus k_A^{n}=1$& $k_B^{m}\oplus k_B^{n}=1$\\
\hline
1& 0& 1& 0& 0& SAME& $k_A^{m}\oplus k_A^{n}=1$& $k_B^{m}\oplus k_B^{n}=1$\\
\hline
1& 0& 1& 1& 1& DIFF& $k_A^{m}\oplus k_A^{n}=1$& $!(k_B^{m}\oplus k_B^{n})=1$\\
\hline
1& 1& 0& 0& 0& SAME& $k_A^{m}\oplus k_A^{n}=0$& $k_B^{m}\oplus k_B^{n}=0$\\
\hline
1& 1& 0& 1& 1& DIFF& $k_A^{m}\oplus k_A^{n}=0$& $!(k_B^{m}\oplus k_B^{n})=0$\\
\hline
1& 1& 1& 0& 1& DIFF& $k_A^{m}\oplus k_A^{n}=0$& $!(k_B^{m}\oplus k_B^{n})=0$\\
\hline
1& 1& 1& 1& 0& SAME& $k_A^{m}\oplus k_A^{n}=0$& $k_B^{m}\oplus k_B^{n}=0$\\
\hline
\end{tabular}
\label{TABLE A}
\end{table}

Here the detector results and the sifted key for Alice(Bob) when $\left |\phi_A^{m}+\phi_B^{n}-\phi_A^{n}-\phi_B^{m}\right |=0$ are listed in Table App. B. The detector results are divided into two types, denoted as SAME which represents the two clicks are triggered by the same detector, and DIFF which represents the two clicks are triggered by different detectors.

\end{document}